\documentclass[aip,superscriptaddress,reprint]{revtex4-1}
\usepackage[english]{babel}
\usepackage{savesym}
\usepackage{amsmath,amsthm,amssymb}
\usepackage{amsfonts}
\usepackage{xspace}
\savesymbol{setrefcountdefault}
\usepackage{graphicx}

\usepackage[separate-uncertainty=true]{siunitx}
\DeclareSIUnit[number-unit-product = {}]
\decade{\text{dec}}
\usepackage{dcolumn}
\usepackage{lmodern}

\usepackage{color}

\begin{document}

\title{Time resolved spin Seebeck effect experiments}

\author{Niklas Roschewsky}
\email{roschewsky@berkeley.edu}
\affiliation{Walther-Mei\ss ner-Institut, Bayerische Akademie der Wissenschaften, Garching, Germany}

\author{Michael Schreier}
\affiliation{Walther-Mei\ss ner-Institut, Bayerische Akademie der Wissenschaften, Garching, Germany}

\author{Akashdeep Kamra}
\affiliation{Walther-Mei\ss ner-Institut, Bayerische Akademie der Wissenschaften, Garching, Germany}
\affiliation{Kavli Institute of Nanoscience, Delft University of Technology, Delft, The Netherlands}

\author{Felix Schade}
\affiliation{Walther-Mei\ss ner-Institut, Bayerische Akademie der Wissenschaften, Garching, Germany}

\author{Kathrin Ganzhorn}
\affiliation{Walther-Mei\ss ner-Institut, Bayerische Akademie der Wissenschaften, Garching, Germany}

\author{Sibylle Meyer}
\affiliation{Walther-Mei\ss ner-Institut, Bayerische Akademie der Wissenschaften, Garching, Germany}

\author{Hans Huebl}
\affiliation{Walther-Mei\ss ner-Institut, Bayerische Akademie der Wissenschaften, Garching, Germany}
\affiliation{Nanosystems Initiative Munich (NIM), Munich, Germany}

\author{Stephan Gepr\"ags}
\affiliation{Walther-Mei\ss ner-Institut, Bayerische Akademie der Wissenschaften, Garching, Germany}

\author{Rudolf Gross}
\affiliation{Walther-Mei\ss ner-Institut, Bayerische Akademie der Wissenschaften, Garching, Germany}
\affiliation{Physik-Department, Technische Universit\"at M\"unchen, Garching, Germany}
\affiliation{Nanosystems Initiative Munich (NIM), Munich, Germany}

\author{Sebastian T.~B. Goennenwein}
\affiliation{Walther-Mei\ss ner-Institut, Bayerische Akademie der Wissenschaften, Garching, Germany}
\affiliation{Nanosystems Initiative Munich (NIM), Munich, Germany}

\date{\today}
\begin{abstract}
In this letter we present the results of transient thermopower experiments, performed at room temperature on yttrium iron garnet/platinum bilayers. Upon application of a time-varying thermal gradient, we observe a characteristic low-pass frequency response of the ensuing thermopower voltage with cutoff frequencies of up to $\SI{37}{\mega\hertz}$. We interpret our results in terms of the spin Seebeck effect, and argue that small wavevector magnons are of minor importance for the spin Seebeck effect in our thin film hybrid structures.
\end{abstract}

\maketitle
{The possibility to drive pure spin angular momentum currents across ferromagnet/metal interfaces by means of thermal gradients -- referred to as ``spin Seebeck effect'' in the literature~\cite{Bauer2012} -- is a fascinating spin-caloritronic concept. However, the microscopic mechanisms and interactions relevant  for spin Seebeck physics still are controversially discussed.~\cite{Adachi2011a,Adachi2013,Tikhonov2013,Chotorlishvili2013,Xiao2010b, Hoffman2013a} To date, most of the spin Seebeck experiments published in the literature were performed using static (DC) or low-frequency thermal gradient drives.~\cite{Weiler2012,Uchida2010d,Agrawal2013,Qu2013} In contrast, time-resolved measurements could allow identifying the time constants relevant for the spin Seebeck effect, thus providing important information about the microscopic mechanisms and interactions at work. In the theoretical model by Xiao \textit{et al.},~\cite{Xiao2010b} magnon-phonon thermalisation time ($\tau_\text{mp}$) influences the spin Seebeck effect. While calculating numbers, Xiao \textit{et al.} further makes the assumption that $\tau_\text{mp}$ is the same as the relaxation time of the uniform precession mode. The latter is then calculated in terms of the Gilbert damping constant. More specifically, according to Xiao \textit{et al.}, in conjunction with their assumption of equality between the magnon-phonon thermalization time ($\tau_\text{mp}$) and the uniform precession mode relaxation time,~\cite{Xiao2010b} one would anticipate a roll-off of the dynamical spin Seebeck effect response in yttrium iron garnet/platinum (YIG/Pt) hybrid structures, if the exciting thermal gradient is modulated with frequencies of about $\SI{1}{\mega\hertz}$. 
In bulk YIG crystals, $\tau_\text{mp}$ for $k=0$ magnons is known to be of the order of several $\SI{100}{\nano\second}$,~\cite{Spencer1962} which results in the roll-off frequencies $f_\text{mod}<1/(2 \pi \SI{100}{\nano\second})=\SI{1.5}{\mega\hertz}$ mentioned above. Thus, transient thermopower (i.e. spin Seebeck) measurements in YIG/Pt hybrids  are interesting.\\ }
In this letter, we present transient {thermopower measurements in YIG/Pt thin film bilayers}, showing no change of the {thermopower} voltage up to frequencies of about $\SI{30}{\mega Hz}$. The measurements are performed in the {so-called longitudinal spin Seebeck} configuration,~\cite{Uchida2010d} using a focused laser beam to generate a temperature gradient along the YIG/Pt hybrid surface normal. Although the laser heating also induces in-plane thermal gradients, corresponding contributions effectively cancel owing to the radial symmetry of the in-plane gradient.
As consequence of the laser heating we observe a thermal voltage which we, in accordance with previously published results,~\cite{Weiler2012,Uchida2010d,Agrawal2013,Qu2013} identify as being caused by the spin Seebeck effect, owing to its symmetry with respect to an externally applied magnetic field and its linear scaling with the thermal gradient magnitude.
The laser directly deposits heat in the electron system of the metal, from which it is distributed to the phonon and magnon systems. Steady state is reached after a {time $t\approx\tau^*$}, the slowest relevant time constant in the system. Our time resolved spin Seebeck measurements are performed with an effective bandwidth of $\SI{50}{\mega\hertz}$ of the detection electronics, but are ultimately limited to a bandwidth of $\lesssim\SI{30}{\mega\hertz}$ by the characteristic low-pass behavior of the samples. Since we observe no evidence for an \textit{intrinsic} process limiting the {thermopower response} within our measurement bandwidth we conclude that {$\tau^*<2\pi\SI{30}{\mega\hertz}\equiv\SI{5}{\nano\second}$.} {This implies that the interaction of phonons with small wavenumber magnons with $\tau_{mp}\gtrsim\SI{100}{\nano\second}$ cannot be a relevant mechanism for the spin Seebeck effect in our samples.
\\}
\begin{figure}
  \includegraphics[bb = 0 0 241 184]{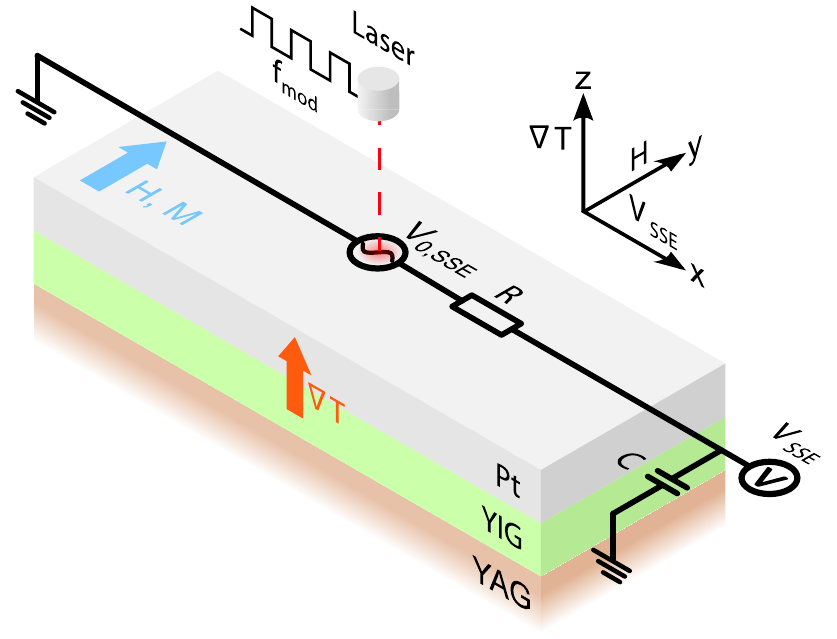}
  \caption{An intensity modulated red laser generates a temperature gradient $\nabla T$ along the YIG/Pt surface normal. The ensuing spin current across the YIG/Pt interface is converted to a charge current by the inverse spin Hall effect. For open circuit conditions a voltage drop along the Pt layer is measured. The polarization vector of the spin current is controlled by an external magnetic field $H$ applied in $y$-direction, perpendicular to both $\nabla T$ and the measurement direction $x$. The voltage detection is ultimately limited by the low-pass behavior of the sample and the measurement electronics, which are sketched as an equivalent circuit diagram.}
  \label{fig:Setup}
\end{figure}
The samples used for the experiments are YIG/Pt bilayers deposited onto single crystalline, ($111$) oriented yttrium aluminium garnet substrates ($\text Y_3\text{Al}_5\text O_{12}$, YAG). The epitaxial YIG film is grown via pulsed laser deposition, and the Pt is deposited in-situ, without breaking the vacuum, on top of the YIG by electron beam evaporation (for details on the sample fabrication see Ref.~\onlinecite{Geprags2012a}). In the following, we will refer to the individual samples as YIG($a$)/Pt($b$), the numbers $a$ and $b$ in brackets indicating the individual layer thicknesses in nanometers determined by high resolution x-ray diffraction. Using optical lithography and argon ion milling, the films are patterned into rectangular shaped stripes.\\
The measurement setup is sketched in Fig.~\ref{fig:Setup}. We use an electrically modulated red laser (\textit{Toptica iBeam smart}, $\lambda_\text{Laser}=\SI{645}{\nano\metre}$) coupled to an optical fiber which terminates in a collimation pack focusing the laser beam to a small spot of about $\SI{2.5}{\micro m}$ diameter.~\cite{Weiler2012, Schreier2013} The focused laser beam locally heats the sample, thereby creating a thermal gradient across the YIG/Pt interface (parallel to the surface normal). For spatially resolved measurements the laser spot can be scanned across the sample surface by means of a xyz-stage. All data discussed in this letter was taken at room temperature. To distinguish the spin Seebeck effect from other thermal effects we exploit the fact that the spin Seebeck voltage ($V_\mathrm{SSE}$) depends on the applied external magnetic field direction via the inverse spin Hall effect, which converts the spin current $\boldsymbol{J_s}$ to a charge current $\boldsymbol{J_c}$~\cite{Tserkovnyak2002}:
\begin{equation}
  \boldsymbol{J}_c= -\theta_\mathrm{H}\frac{2e}{\hbar}\boldsymbol{J}_s\times \boldsymbol\sigma.\label{eq:ISHE}
\end{equation}
Here, $\boldsymbol\sigma$ is the spin polarization determined by an externally applied field parallel to the YIG film along the $y$-direction, i.e. perpendicular to the thermal gradient $\boldsymbol \nabla  T$ (cf. Fig.~\ref{fig:Setup}) and $\theta_\mathrm{H}$ is the spin Hall angle.~\cite{Tserkovnyak2002} For simplicity, we refer to our experimental data as spin Seebeck effect, although strictly speaking we only expect a thermopower consistent with~\eqref{eq:ISHE}.
\begin{figure}
  \includegraphics[bb=0 0 241 171]{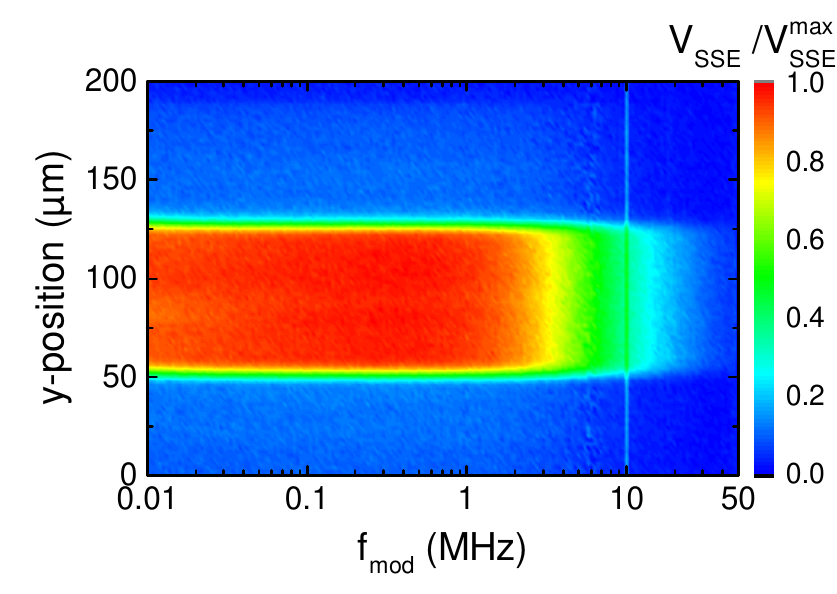}
  \caption{False color plot of the normalized spin Seebeck voltage for the YIG(48)/Pt(4.9) sample. At a fixed frequency $f_\mathrm{mod}$, the laser scans the sample in $\SI{2}{\micro m}$ steps in the $y$-direction, and the voltage along the Pt bar at every point is measured. This procedure is then repeated for the next frequency. The $V_\mathrm{SSE}$ data shown here are obtained by subtracting two sets of data recorded at $+\SI{70}{\milli T}$ and $-\SI{70}{\milli T}$, respectively. The line at $f_\mathrm{mod}\approx\SI{10}{\mega\hertz}$ is an electrical resonance in the detection electronics circuit. The Pt stripe extends between $y=\SI{50}{\micro\metre}$ and $\SI{130}{\micro\metre}$ along the $y$-axis. On this sample the spin Seebeck voltage signal stays constant up to about $f=\SI{1}{\mega\hertz}$ and then decays with increasing $f_\mathrm{mod}$ due to the RC low pass filtering.}
  \label{fig:FrequencyScan}
\end{figure}
In our measurements, we use a square waveform modulation of the laser intensity at frequency $f_\mathrm{mod}$, resulting in a periodic heating at the same frequency. Then, via the spin Seebeck effect, the sample acts as a voltage source with a square wave output of amplitude $V_{0,\mathrm{SSE}}$. Due to the finite resistivity of the Pt film the sample acts as a load resistor $R$ to the voltage source. The sample is connected to the measurement electronics via coaxial cables, which, in addition to the sample carrier, grounds $V_{0,\mathrm{SSE}}$ via a shunt capacitance $C$. The equivalent circuit diagram is shown in Figure~\ref{fig:Setup}. Using Kirchhoff`s law the frequency response of the system can be modeled as
\begin{equation}
  RC \frac{\text d V_\text{SSE}}{\text d t}+ V_\text{SSE}=V_{0,\mathrm{SSE}}.\label{eq:lowpass}
\end{equation}
where $V_\mathrm{SSE}$ is the voltage measured across the Pt and $V_{0,\mathrm{SSE}}$ is the ``true" voltage generated by the spin Seebeck effect. Equation~\eqref{eq:lowpass} yields the characteristic first order low-pass transfer function 
\begin{equation}
	V_\text{SSE}(f_\mathrm{mod})=\frac{V_{0,\mathrm{SSE}}}{\sqrt{1+(f_\mathrm{mod}/f_\text{c})^2}}
\label{eq:lowpasstransfer}
\end{equation}
with the cutoff frequency $f_\text{c}=(2\pi RC)^{-1}$. For modulation frequencies $f_\mathrm{mod}\ll f_\mathrm{c}$, one thus will observe the full spin Seebeck voltage $V_{0,\mathrm{SSE}}$, while for $f_\mathrm{mod}>f_\mathrm{c}$ the $V_\mathrm{SSE}$ observed in experiment should show a characteristic RC-filter $1/f_\mathrm{mod}$ type roll-off.\\
Figure~\ref{fig:FrequencyScan} shows a spatially and frequency resolved spin Seebeck measurement. Here, each vertical line represents a spatial scan of the laser spot in the $y$-direction, across the Pt bar, for fixed laser modulation frequency $f_\mathrm{mod}$. We used a \textit{Zurich Instruments HF2LI} lock-in amplifier to record $V_\mathrm{SSE}$. Whenever the laser spot is on the Pt bar ($\SI{50}{\micro m}\leq y\leq\SI{130}{\micro m}$), a large voltage ($\sim\SI{1}{\micro V}$) is detected that inverts its sign on reversing the direction of the applied magnetic field, as expected for the spin Seebeck effect viz. the spin Hall effect. Note that we here discuss the spin Seebeck voltage signal obtained by subtracting two measurements with opposite magnetic field orientation: $V_\mathrm{SSE}=V_\mathrm{SSE,raw}(+H)-V_\mathrm{SSE,raw}(-H)$ with $|\mu_0H|=\SI{70}{\milli T}$. Additionally, the data in Fig.~\ref{fig:FrequencyScan} are normalized to the maximum measured spin Seebeck voltage $V_\mathrm{SSE}^\mathrm{max}$ for clarity. In the YIG(48)/Pt(4.9) sample shown in Fig.~\ref{fig:FrequencyScan}, $V_\mathrm{SSE}$ is essentially constant up to a modulation frequency of a few megahertz and decays with $\SI{20}{\dB/\decade}$ at higher frequencies. As noted above, this decay is due to the RC limitation of our measurement setup rather than intrinsic spin Seebeck physics.\\
\begin{figure}
  \includegraphics[bb=0 0 241 171]{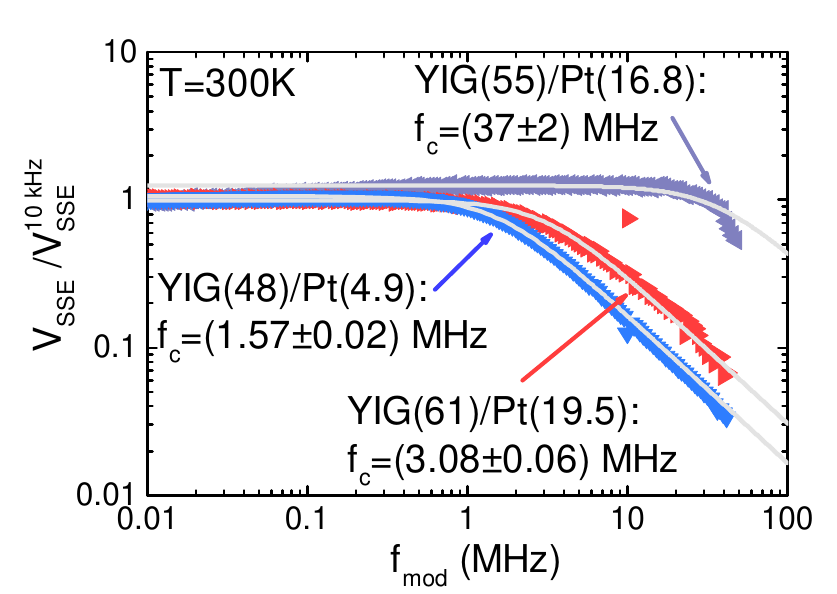}
  \caption{Normalized spin Seebeck voltage as a function of the laser modulation frequency $f_\mathrm{mod}$ at a fixed position at the center of the Pt bar for three different samples. In all samples $V_\mathrm{SSE}$ is essentially frequency independent up to about \SI{1}{\mega\hertz}. The roll-off in $V_\mathrm{SSE}$ observed for $f_\mathrm{mod}\gtrsim\SI{1}{\mega\hertz}$ is quantitatively described by the low-pass transfer function [Eq.~\protect\eqref{eq:lowpasstransfer}], using a shunt capacitance of $C=\SI{125\pm39}{\pico F}$ and the resistance $R$ of the respective samples. The corresponding $\SI{3}{\dB}$ low-pass cutoff frequency $f_\mathrm{c}$ is quoted for each sample.}
  \label{fig:fvsSSE}
\end{figure}
To address the low-pass behavior of our measurement circuit in more detail, we measured the spin Seebeck voltage at one fixed ($x,y$) position in the center of the Pt bar as a function of the laser modulation frequency for three different samples. As evident from Fig.~\ref{fig:fvsSSE}, the YIG(48)/Pt(4.9) sample ($R=\SI{876}{\ohm}$) shows the lowest $\SI{3}{\dB}$ cutoff frequency $f_\mathrm{c}$ of about $\SI{1.57}{\mega\hertz}$, while the YIG(61)/Pt(19.5) sample ($R=\SI{309}{\ohm}$) exhibits a slightly larger one ($\SI{3.08}{\mega\hertz}$). The third sample [YIG(55)/Pt(16.8)] was designed to exhibit a small resistance ($R=\SI{47}{\ohm}$) which results in a $\SI{3}{\dB}$ cutoff frequency of about $\SI{37}{\mega\hertz}$. Since the lock-in amplifier used in the experiments has a bandwidth limited to $\SI{50}{\mega\hertz}$, $V_\mathrm{SSE}$ decays faster than the $\SI{20}{\dB/\decade}$ expected from the first order RC low-pass for frequencies close to $\SI{50}{\mega\hertz}$. Note that all data in Fig.~\ref{fig:fvsSSE} can be fitted with Eq.~\eqref{eq:lowpasstransfer} with only a single fitting parameter ($f_\mathrm{c}$). From the relation $f_\text{c}=(2\pi RC)^{-1}$ we can then extract the common shunt capacitance $C=\SI{125\pm39}{\pico F}$ of our measurement setup.\\ 
\begin{figure}
  \includegraphics[bb=0 0 241 256]{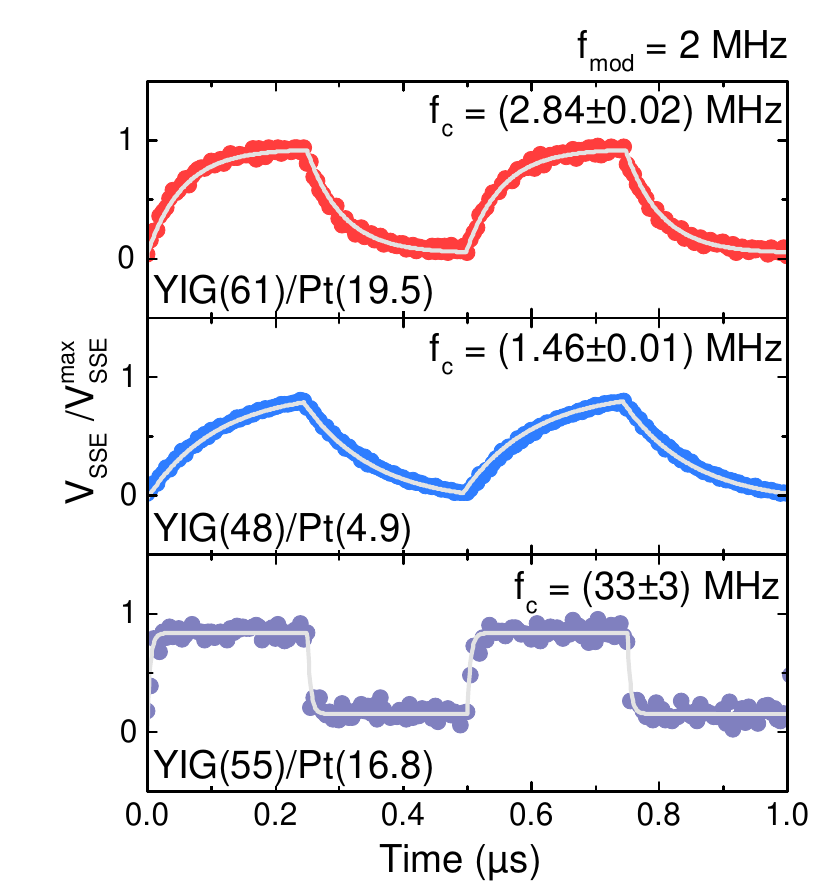}
  \caption{Spin Seebeck voltage as a function of time for laser modulation frequencies of $\SI{2}{\mega\hertz}$ for the three investigated samples. The two samples with large resistance show a shark fin like response to the excitation, while the voltage response of the third, low resistance sample [YIG(55)/Pt(16.8)] is still rectangular. This behavior is well reproduced by the fitted low-pass response, indicated by the solid lines, from which the $\SI{3}{\dB}$ cutoff frequency is obtained. }
  \label{fig:TimeTrace}
\end{figure}
In addition to the frequency dependent measurements we performed measurements in the time domain. 
For these measurements we used a \textit{GaGe Razor} digitizer card and a \textit{FEMTO HVA-200M-40-F} [YIG(48)/Pt(4.9) and YIG(61)/Pt(19.5)] or \textit{FEMTO DHPVA-200} [YIG(55)/Pt(16.8)] preamplifier to record the voltage arising from the laser heating. The different amplifiers were chosen to avoid electrical resonances in the circuit at the expense of different signal-to-noise ratios. At $f_\mathrm{mod}=\SI{2}{\mega\hertz}$ [Fig.~\ref{fig:TimeTrace}] the two samples with large resistance show a shark fin like response to the laser heating while the response of the third sample is still square wave shaped. The time traces can be fitted with the transient solution to Eq.~\eqref{eq:lowpass} which then again yields the individual cutoff frequencies. The small difference between the values for $f_\mathrm{c}$ obtained from the lock-in and the time-domain measurements is due to the introduction of additional shunt capacitances in the latter measurement scheme. The behavior of the individual samples is well reproduced by the low-pass model with a common capacitance $C$ and the resistance of the respective sample, corroborating the notion that the observed roll-off of the spin Seebeck voltage at higher modulation frequencies is not arising from intrinsic spin Seebeck physics, but rather due to the limited bandwidth of the experiment.\\
We now interpret our experimental results. {No intrinsic frequency dependence of the thermopower (viz. spin Seebeck) voltage up to at least $\SI{30}{\mega\hertz}$ was observed in our experiments. Thus, we conclude that relevant time constants $\tau^*$ for this caloritronic effect must be smaller than about $\tau=(2\pi f)^{-1}\approx\SI{5}{\nano s}$.}\\
{Assuming the magnon phonon interaction time in our YIG thin films is comparable to the bulk value, this especially rules out small wavenumber magnons and the uniform precession mode $(k=0)$ as the dominant source of the spin Seebeck effect spin current, since their interaction times with phonons are of the order of several hundred nanoseconds.~\cite{Spencer1962}\\}
Indeed, recent theoretical considerations~\cite{Schreier2013, Hoffman2013a} suggest that thermal (large $k$) and not the $k=0$ magnons are responsible for the spin Seebeck effect. {This would be in agreement with the experiments, presented here, since thermal magnons are predicted~\cite{Schreier2013,Hoffman2013a,Cherepanov1993} to thermalize with phonons much faster than their $k=0$ counterparts. However, dedicated experiments are required to fully verify these conjectures.}\\
Furthermore, in addition to $\tau_\mathrm{mp}$, the magnon-magnon interaction time $\tau_\mathrm{mm}$ also could be relevant for the spin Seebeck effect.~\cite{Xiao2010b} The few available data and estimates for this quantity vary over a broad range of about $10^{-7}-\SI{e-9}{s}$.~\cite{Demokritov2006, Kittel1953} Present spin Seebeck theories~\cite{Xiao2010b, Hoffman2013a} assume $\tau_\mathrm{mm}$ to be much smaller than $\tau_\mathrm{mp}$. The time constant of $\SI{5}{\nano s}$, extracted as an upper bound for $\tau_\mathrm{mp}$ from our experiments, thus suggests that $\tau_\mathrm{mm}$ in YIG is closer or even below to the lower end of the above range, at least for the specific subset of magnons responsible for the spin Seebeck effect.\\
Finally, Agrawal \textit{et al.}~\cite{Agrawal2013} recently performed similar experiments on a $\SI{6.7}{\micro m}$ thick YIG sample. From their data they extract a time constant of $\SI{343}{\nano s}$ as an \textit{intrinsic}, characteristic  limit to spin Seebeck physics in YIG due to a finite magnon diffusion length, {based on a simulation of the phonon temperature evolution in their sample.} This time constant may not be immediately transferable to our experiments in thin film samples, in which the respective time constant, as per their explanation, could be shorter than our detection bandwidth. Nevertheless, the experiments on our samples clearly show no such limit for times larger than about $\SI{5}{\nano s}$.\\
In conclusion we experimentally studied transient thermopower voltages arising in YIG/Pt thin film samples upon irradiation with an intensity modulated laser beam. We interpret this as spin Seebeck effect. From both, lock-in measurements of $V_\mathrm{SSE}$ for different modulation frequencies as well as time resolved $V_\mathrm{SSE}$ measurements we consistently find no evidence for a decay of the spin Seebeck signal due to \textit{intrinsic} time constants up to at least $\SI{30}{\mega\hertz}$. Rather, the observed frequency dependent roll off in $V_\mathrm{SSE}$ can be quantitatively understood as a simple RC low-pass filter effect of the experimental setup. 
Our results on time scales up to about $\SI{5}{\nano s}$ (corresponding to the experimental bandwidth of $\SI{30}{\mega\hertz}$) thus truly go beyond~\cite{Bieren2013} Weiler \textit{et al.}~\cite{Weiler2012} and support the notion that small $k$ magnons do not contribute significantly to the spin Seebeck at room temperature.~\cite{Schreier2013, Hoffman2013a} If pumped spin current in the spin Seebeck effect is dominated by thermal magnons [small $\tau_\mathrm{mp}$ (Ref.~\onlinecite{Bhandari1966})], one can expect the spin Seebeck effect to be robust up to much higher frequencies ($\SI{}{\giga\hertz}$ and above). While {this notion} hinges on the assumption that {the magnon-phonon interaction is indeed the limiting time constant in the spin Seebeck effect}, a notion which is supported by many experiments but for which direct proof is still missing, our data show that the mechanism behind the spin Seebeck effect is robust with respect to thermal excitation frequencies of the order of at least several ten megahertz.\\
\\
We would like to thank Matthias Opel for valuable discussions and gratefully acknowledge financial support from the DFG via SPP 1538 ``Spin Caloric Transport'' (project GO 944/4-1).

\bibliography{Time_resolved_spin_Seebeck_effect}

\end{document}